\documentclass{jpp}

\newcommand{\apjl}{\mbox{\it Astrophysical Journal Letters}}
\newcommand{\apj}{\mbox{\it Astrophysical Journal}}

\newcommand{\mnras}{\mbox{\it Mon. Not. R. Astron. Soc.}}

\newcommand{\pre}{\mbox{\it Phys. Rev. E}}

\usepackage{color}
\usepackage{graphicx}
\usepackage{natbib}

\ifCUPmtlplainloaded \else
  \checkfont{eurm10}
  \iffontfound
    \IfFileExists{upmath.sty}
      {\typeout{^^JFound AMS Euler Roman fonts on the system,
                   using the 'upmath' package.^^J}%
       \usepackage{upmath}}
      {\typeout{^^JFound AMS Euler Roman fonts on the system, but you
                   dont seem to have the}%
       \typeout{'upmath' package installed. JPP.cls can take advantage
                 of these fonts, if you use 'upmath' package.^^J}%
      }
  \else
  \fi
\fi

\ifCUPmtlplainloaded \else
  \checkfont{msam10}
  \iffontfound
    \IfFileExists{amssymb.sty}
      {\typeout{^^JFound AMS Symbol fonts on the system, using the
                'amssymb' package.^^J}%
       \usepackage{amssymb}%

      }{}
  \fi
\fi

\ifCUPmtlplainloaded \else
  \IfFileExists{amsbsy.sty}
    {\typeout{^^JFound the 'amsbsy' package on the system, using it.^^J}%
     \usepackage{amsbsy}}
    {}
\fi

\newsavebox{\astrutbox}
\sbox{\astrutbox}{\rule[-5pt]{0pt}{20pt}}

\title[Density jump for oblique collisionless shocks in pair plasmas: physical solutions]{Density jump for oblique collisionless shocks in pair plasmas: physical solutions}

\author[A. Bret and Colby C. Haggerty and R. Narayan]%
{Antoine Bret$^{1,2,3,4}$, Colby C. Haggerty$^5$, Ramesh Narayan$^{3,4}$%
  \thanks{Email address for correspondence: antoineclaude.bret@uclm.es}
}

\affiliation{$^1$ETSI Industriales, Universidad de Castilla-La Mancha, 13071 Ciudad Real, Spain\\[\affilskip]
$^2$Instituto de Investigaciones Energ\'{e}ticas y Aplicaciones Industriales, Campus Universitario de Ciudad Real, 13071 Ciudad Real, Spain\\[\affilskip]
$^3$ Center for Astrophysics | Harvard \& Smithsonian, Harvard University, 60 Garden St., Cambridge, MA 02138, USA\\
$^4$Black Hole Initiative at Harvard University, 20 Garden St., Cambridge, MA 02138, USA\\
$^5$Institute for Astronomy, University of Hawaii, Manoa, 2680 Woodlawn Dr., Honolulu, HI 96822, USA
}

\date{?; revised ?; accepted ?. - To be entered by editorial office}

\begin{document}

\maketitle

\begin{abstract}
Collisionless shocks are frequently analyzed using the magnetohydrodynamics (MHD) formalism, even though MHD assumes a small mean free path. Yet, isotropy of pressure, fruit of binary collisions and assumed in MHD, may not apply in collisionless shocks. This is especially true within a magnetized plasma, where the field can stabilize an anisotropy.

 In a previous article \citep{BretJPP2022b}, a model was presented capable of dealing with the anisotropies that may arise at the front crossing. It was solved for any orientation of the field with respect to the shock front. Yet, for some values of the upstream parameters, several downstream solutions were found.

 Here, we complete the work started in \cite{BretJPP2022b} by showing how to pick the physical solution out of the ones offered by the algebra. This is achieved by 2 means: 1) selecting the solution that has the downstream field obliquity closest to the upstream one. This criterion  is exemplified on the parallel case and backed up by Particle-in-Cell simulations. 2) Filtering out solutions which do not satisfy a criteria already invoked to trim multiple solutions in MHD: the evolutionarity criterion, that we assume valid in the collisionless case.

 The end result is a model in which a given upstream configuration results in a unique, or none (like in MHD), downstream configuration. The largest departure from MHD is found for the case of a parallel shock.
\end{abstract}

\maketitle

\section{Introduction}
Shock waves are fundamental processes in fluids. They have been the subject of numerous studies for nearly two centuries \citep{Salas2007}. When the frequency of collisions between particles is high, the thickness of the shock front is of the order of a few mean free paths, since binary collisions are ultimately the only microscopic mechanism capable of transferring some of the kinetic energy of the upstream medium into heat in the downstream \citep{Zeldovich}.

However, \emph{in situ} observations of the bow shock of the Earth's magnetosphere in the solar wind have shown that its front is about a hundred kilometers thick, while the mean free path at the same location is of the order of the Sun-Earth distance \citep{PRLBow1,PRLBow2}. Such a shock cannot be mediated by collisions. It is mediated by collective plasma electromagnetic effects \citep{Sagdeev66}. This type of shock is known as ``collisionless shock''.

Strictly speaking, collisionless shocks should be studied at the kinetic level, using the Vlasov equation, since the absence of collisions can even make it difficult to define a local velocity, as is the case, for example, in counter streaming systems. Due to the complexity involved in solving the Vlasov equation, collisionless shocks, and in particular the density, temperature or velocity jumps they present, are often interpreted via MHD.

Yet MHD relies on hydrodynamics and as such entails the same hypothesis of small mean free path \citep{gurnett2005,Goedbloed2010,TB2017}. This hypothesis of small mean free path has 2 consequences that are important for the calculation of the density jump around a shock. The first consequence is that pressure is isotropic, both before and after the shock. In fact, even if a fluid is subjected to pressure anisotropy during its transport from the upstream to the downstream, binary collisions will restore isotropy on a time scale of the order of the collision frequency, well below the macroscopic times involved. The second consequence is that all of the upstream fluid passes into the downstream, along with the matter and momentum it carries.

It turns out that in a collisionless shock, these 2 consequences can be invalidated \citep{BretApJ2020}. The first, because in the absence of collisions, a plasma can maintain a stable anisotropy in the presence of an ambient magnetic field \citep{Hasegawa1975,Gary1993}. The second consequence, because due to the large mean free path, plasma particles can bounce off the shock front or even travel upstream from the downstream, as is the case with accelerated particles \citep{Drury1983,Blandford1987}.

This article proposes a remedy to the first consequence: how to correct the MHD jump equations so that they can account for an anisotropy in the plasma?

Notably, in the absence of an ambient magnetic field, the Weibel instability ensures isotropy of a collisionless plasma \citep{Weibel,SilvaPRE2021}. Therefore, the MHD jump equations only need to account for anisotropies if a magnetic field is present.

Several authors have derived the MHD jump equations for the non-isotropic case \citep{Hudson1970,Karimabadi95,Erkaev2000,Gerbig2011}. But in all of these works, while the anisotropy of the upstream is considered a free parameter, so is that of the downstream. These equations are therefore unable, on their own, to derive the density jump of a shock whose downstream is not isotropic, because they lack precisely one parameter: the anisotropy of the downstream.

In a recent series of papers, we developed a model that precisely fills this gap. It was first applied to the case of a parallel shock \citep{BretJPP2018}, i.e. a shock moving parallel to the ambient magnetic field. The assumptions made and the results obtained were confirmed by numerical simulations \citep{Haggerty2022}.

The model was then applied to the case of a perpendicular shock \citep{BretPoP2019}, and finally to the case of a switch-on and of an oblique shock \citep{BretJPP2022a,BretJPP2022b}. The latter case, that of an oblique shock, is analytically much more complicated than the parallel and perpendicular cases, due to the complexity of the MHD jump equations for an oblique field and anisotropic temperatures.

In \cite{BretJPP2022b}, hereafter referred to as Paper I, the algebra of these equations was solved, but the solutions were left unfiltered. As a result, several coexisted for some combinations of the upstream parameters.

Here, Paper I is completed by filtering the algebraic solutions so that for a given combination of upstream variables, there is no more than 1 solution for the downstream.

\begin{figure}
\begin{center}
 \includegraphics[width=\textwidth]{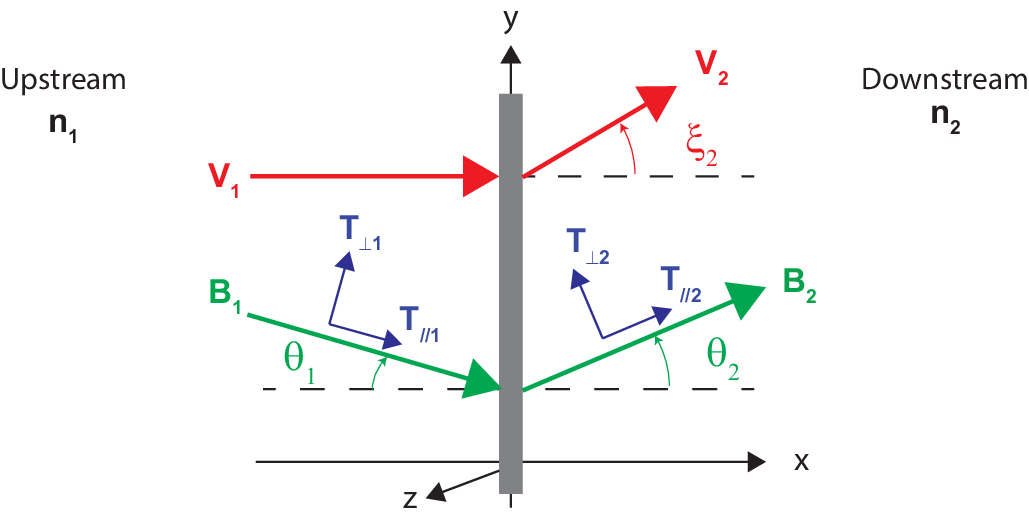}
\end{center}
\caption{System under scrutiny. It is identical to that of Paper I. Although the conservation equations used are valid for any upstream temperatures, we here, like in Paper I, only treat the strong sonic case $T_{\perp 1}=T_{\parallel 1}=0$. We work in the frame of reference where $\mathbf{v}_1$ is normal to the front.}\label{fig:system}
\end{figure}

\section{Method}
The system under scrutiny is represented on  Figure \ref{fig:system}. It is identical to that of Paper I. Though the conservation equations we shall use are valid for any upstream temperatures, we here, like in Paper I, only treat the strong sonic case $T_{\perp 1}=T_{\parallel 1}=0$.

\subsection{Summary of previous works}
As previously said, the basic caveat of MHD is that if a collisionlessly stable anisotropy develops at the front crossing, MHD itself cannot derive it. The jump of quantities like the density is therefore under-determined.

For completeness, we now briefly recall the results obtained in previous works.

In \cite{BretJPP2018} we presented a model capable of solving this issue for a parallel shock. We reasoned that as it crosses the front, the plasma is compressed in the direction parallel to the motion. As a consequence, its parallel temperature increases while its perpendicular temperature remains constant. The 3 MHD conservation equations (matter, momentum, energy\footnote{For a parallel shock, the MHD conservation equations are identical to the fluid ones \citep{Kulsrud2005}. With anisotropic temperature, they are obtained setting $\theta_1=\theta_2=\xi_2=0$ in Eqs. (\ref{eq:conser1}-\ref{eq:conser6}) of Appendix \ref{ap:MHDoblique}.}) are therefore completed by,
\begin{equation}\label{eq:Tperp}
T_{\perp 2} = T_{\perp 1},
\end{equation}
allowing to derive the 4 downstream unknowns ($n_2,v_2,T_{\perp 2},T_{\parallel 2}$), in terms of the upstream variables. Note that here, the perpendicular direction is common to the flow and the field.

Now, the state of the downstream resulting from the conservation of $T_\perp$ may be stable, or not. If it is stable, then this is the end state of the downstream. If it is unstable, the plasma migrates to the instability threshold\footnote{The nature of this instability will be specified shortly.}. Imposing marginal stability then provides again a fourth equation allowing to fully determine the properties of the downstream.

\cite{BretJPP2018}, as well as every subsequent works of ours on the same model, is limited to pair plasmas. The reason for this is that the equality of the mass of the species allows to consider only one parallel and one perpendicular temperature. In an electron/ion plasma where electrons and ions are heated differently in the shock, a 4 temperatures model would be required \citep{Guo2017,Guo2018}. Yet, since the model eventually relies on macroscopic physics, it should also apply to electron/ion plasmas, as preliminary Particle-in-Cell (PIC) simulations seem to indicate \citep{Shalaby2022}.

The model predicted, for a strong sonic parallel shock, a density jump going from 4 to 2 in the high field regime, a prediction successfully confirmed by PIC simulations in \cite{Haggerty2022}. Such a result stands in  contrast with MHD where the density jump should always be 4, regardless of the field strength\footnote{The jump of a strong fluid shock with adiabatic index $\gamma=5/3$ is 4. And for the parallel case, the MHD and fluid conservation equations are identical \citep{Kulsrud2005}.}.

The perpendicular case was treated in \cite{BretPoP2019}. There, the direction perpendicular to the flow is eventually parallel to the field, so that the counterpart of Eq. (\ref{eq:Tperp}) is,
\begin{equation}\label{eq:Tpara}
T_{\parallel 2} = T_{\parallel 1}.
\end{equation}

The switch-on shocks, where the field is oblique in the  downstream only, was treated in \cite{BretJPP2022a}. The model has also been solved for a parallel or a perpendicular shock, with an anisotropic \emph{upstream} \citep{BretMNRAS2023a,BretMNRAS2023b}.

Finally, the general case where the field may be oblique in both the upstream and the downstream was treated in \cite{BretJPP2022b}, namely, Paper I.

 In \cite{BretJPP2022a,BretJPP2022b}, the closure of the MHD jump equations was achieved through an \emph{ansatz} interpolating between (\ref{eq:Tperp}) and (\ref{eq:Tpara}). In the limit of a cold upstream with $T_1=0$, which is the regime treated in Paper I and hereafter, the \emph{ansatz} reads,
\begin{eqnarray}\label{eq:ansatz}
  T_{\parallel 2} &=& T_e\cos^2\theta_2, \nonumber\\
  T_{\perp 2}     &=& \frac{1}{2}T_e\sin^2\theta_2,
\end{eqnarray}
where $T_e$ is a parameter determined when solving the equations and $\theta_2$ is the angle of the downstream field with the shock normal (see Figure \ref{fig:system}). Eqs. (\ref{eq:ansatz}) correctly reduce to (\ref{eq:Tperp}) and (\ref{eq:Tpara}) in their respective limits since $\theta_2=0$ for the parallel case,  while $\theta_2=\pi/2$ for the perpendicular one. Such a scheme guarantees that both perpendicular temperatures are equal, as required by the Vlasov equation (\cite{LandauKinetic}, \S 27). Also, the total thermal energy in the 3 directions sums up to $k_BT_e$, where $k_B$ is the Boltzmann constant.

In summary, and since the 2 instabilities involved are the firehose and the mirror instabilities (see Section \ref{sec:insta}), our model can be stated as follow:
\begin{itemize}
  \item As the plasma goes through the shock front, its temperature normal to the flow is conserved for the parallel and perpendicular cases. This translates directly to (\ref{eq:Tperp}) and (\ref{eq:Tpara}) respectively. For the oblique case with cold upstream $T_1=0$, Eqs. (\ref{eq:ansatz}) interpolate between these 2 extremes.
  \item The resulting state of the plasma in the downstream is called ``\textbf{Stage 1}''.
  \item If the downstream field is strong enough to stabilize Stage 1, then this is the end state of the downstream.
  \item If the downstream field is too weak to stabilize Stage 1, then
  \begin{itemize}
    \item If Stage 1 is firehose unstable, it migrates to the firehose instability threshold. This is ``\textbf{Stage 2 - firehose}''.
    \item If Stage 1 is mirror unstable, it migrates to the mirror instability threshold. This is ``\textbf{Stage 2 - mirror}''.
  \end{itemize}
\end{itemize}

\subsection{Present work}
Paper I has  3 kinds of limitations:
\begin{enumerate}
  \item It is restricted to strong sonic shocks, namely $T_1=0$, and to non-relativistic pair plasmas. These  restrictions are still considered here.
  \item It considers the simplest expressions for the Alfv\'{e}n velocity and the stability criterion  of the instabilities involved in our model. Yet, more accurate expressions are required in an anisotropic plasma. The present work accounts for one of them.
  \item It only presents the \emph{allowed} solutions to the conservation equations, plus Eq. (\ref{eq:ansatz}). It does not filter these solutions according to their physical relevance. Such a filtering is the main goal of the present work.
\end{enumerate}

Our purpose here is to deal with points (\emph{b}) and (\emph{c}) above.

Besides the variables explained on Figure \ref{fig:system}, we shall use the following dimensionless parameters,
\begin{eqnarray}\label{eq:dimless}
r &=& \frac{n_2}{n_1},  \nonumber \\
A_i &=& \frac{T_{\perp i}}{T_{\parallel i}},  \nonumber \\
M_{A,i} &=& \frac{v_i}{v_{A,i}}, \nonumber \\
\sigma &=& \frac{B_1^2/8\pi}{\frac{1}{2}n_1 m v_1^2} \equiv \frac{1}{M_{A,1}^2},
\end{eqnarray}
where $v_{A,i}$ is the Alfv\'{e}n velocity,
\begin{equation}\label{eq:Alfviso}
v_{A,i} = \frac{B_i}{\sqrt{4 \pi n_i m}}.
\end{equation}
The parameter $\sigma$ is frequently used in simulations of collisionless shocks like \cite{Haggerty2022}, while the Alfv\'{e}n Mach number is common in MHD shock literature.

In addition to the Alfv\'{e}n Mach number defined above, we shall often use in the sequel its following variant,
\begin{equation}\label{eq:MAx}
M_{Aix} \equiv \frac{v_i \cos\xi_i}{v_{A,i}\cos\theta_i}.
\end{equation}
It compares the projection of the flow velocity along the shock normal ($x$ axis), to the projection of the Alfv\'{e}n velocity, still along the shock normal.

Since the road map for solving our model is eventually the one already used in MHD, we start by reminding the reader how it applies there.

\section{Isotropic MHD results and evolutionarity}\label{sec:MHD}
One of the criteria used here to filter out some solutions is the so-called ``evolutionarity criterion'', already present in isotropic MHD.
It is therefore convenient to present how it operates there.

The MHD conservation equations for isotropic temperatures are reported in Appendix  \ref{ap:MHD}. They can be used to derive an expression of the downstream field angle $\theta_2$ in terms of the upstream quantities only. Namely, the quantity,
\begin{equation}\label{eq:T_2}
T_2 \equiv \tan\theta_2,
\end{equation}
is a root of the polynomial (\ref{eq:T2MHD}) in Appendix \ref{ap:MHD}. The MHD density jump is then given by
\begin{equation}\label{eq:rMHD}
r = \frac{M_{A1}^2 T_2}{M_{A1}^2 \tan \theta_1+T_2 \cos ^2\theta_1-\sin \theta_1 \cos \theta_1},
\end{equation}
in terms of the upstream Alfv\'{e}n Mach number (\ref{eq:dimless}).

Although such a derivation of the MHD density jump is uncommon in the literature, it mimics the derivation of the jump in our model. It is therefore pedagogically valuable for the present work.

\begin{figure}
\begin{center}
 \includegraphics[width=\textwidth]{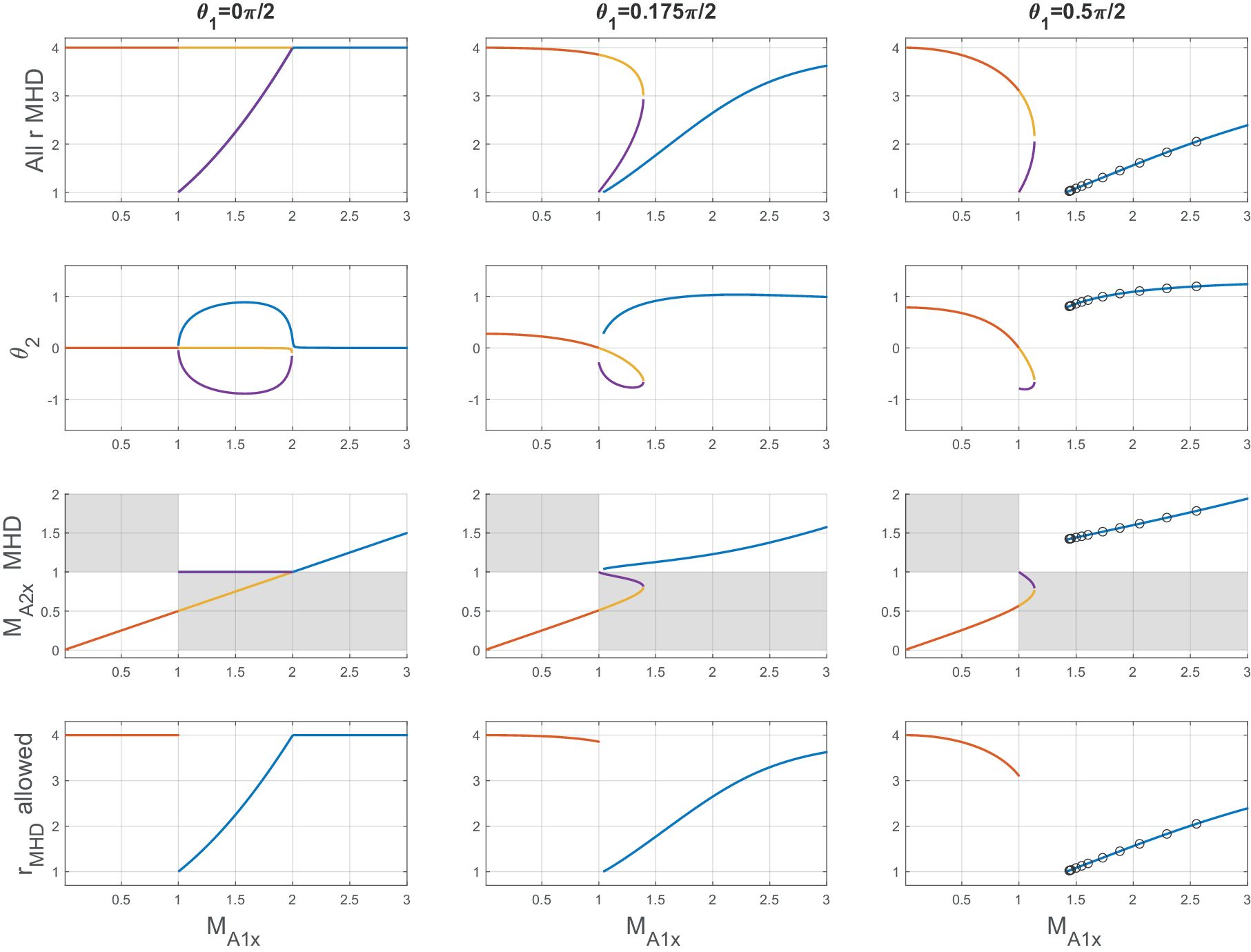}
\end{center}
\caption{\emph{Upper row}: density jump $r$ for different angles $\theta_1$ in terms of $M_{A1x}=M_{A1}/\cos\theta_1$, defined by Eq. (\ref{eq:MAx}). Sometimes there is more than 1 solution. \emph{Second row}: angle $\theta_2$ of the field $\textbf{B}_2$ with the $x$ axis (see Figure \ref{fig:system}). \emph{Third row}: evolutionarity criterion. Some branches, namely, the ones crossing the shaded areas, are to be excluded. \emph{Lower row}: same as upper row, but without the branches excluded by the evolutionarity criterion. There is now but 1 solution for a given $M_{A1x}$, or none. The black circles for $\theta_1=0.5\pi/2$ are the results of our MHD simulations (see Section \ref{eq:MHDsim}).}\label{fig:MHD}
\end{figure}

Figure \ref{fig:MHD} presents the solutions for 3 different angles $\theta_1$. The upper row shows all the possible solutions. For $\theta_1=0$ and $M_{A1x}\in[1,2]$, there are 2 MHD branches, the lower one pertaining to the switch-on solutions. For $\theta_1=0.175 \pi/2$ there are even 3 solutions for $M_{A1x}\in[1, 1.34]$.

Which one will the shock choose? This question is at the heart of this work. Let us now see how it is solved in MHD.

The MHD answer relies on the notion of shock ``evolutionarity'', which has been discussed several times in the literature (e.g., \cite{Kennel1990,Farris1994,FalleJPP2001,Kulsrud2005}  ). For given upstream and downstream boundary conditions, the MHD Rankine-Hugoniot jump conditions give a unique solution for fast shocks, where the flow speed on both sides of the shock are super-Alfv\'{e}nic, and also for slow shocks, where the fluid is throughout sub-Alfv\'{e}nic. These two types of shocks are stable and are described as evolutionary. Four other potential shock types, each of which has super-Alfv\'{e}nic upstream fluid and sub-Alfv\'{e}nic downstream fluid, have no unique solutions to the Rankine-Hugoniot relations. In these shocks, the transverse magnetic field switches sign across the shock, and the fluid equations do not provide the correct number of Alfv\'{e}n waves to handle the field flip (Falle \& Komissarov 1997, Kulsrud 2005). Such shocks will not arise naturally from generic initial conditions. Even if they are artificially set up, they will tend to deviate quickly from their initial configuration, typically splitting into two shocks, one fast and the other slow. These ``forbidden" shocks do \emph{not} satisfy the evolutionarity condition. They are called ``intermediate'' \citep{Komissarov1997} or ``extraneous'' \citep{Kulsrud2005} shocks.

The MHD solutions presented on the upper row of Figure \ref{fig:MHD} need therefore to be filtered according to the aforementioned evolutionarity criterion. The upstream Alfv\'{e}n Mach number to consider for the evolutionarity analysis is not the one given by Eq. (\ref{eq:dimless}), but its variant (\ref{eq:MAx}) instead. With this definition, the downstream Alfv\'{e}n Mach number reads
\begin{equation}\label{eq:MA2MHD}
M_{A2x}^2 = 1+\frac{\tan \theta_1 \left(M_{A1}^2 \sec ^2\theta_1-1\right)}{T_2}.
\end{equation}
It is with definition (\ref{eq:MAx}) that the Alfv\'{e}n Mach number of the MHD switch-on shocks, see left plot of third row on Figure \ref{fig:MHD},  is found equal to 1 (\cite{Goedbloed2010}, p. 853). These modes, by definition, do \emph{not} have $\theta_2=0$ (second row on Figure \ref{fig:MHD}) nor $\xi_2=0$. Corrections (\ref{eq:MAx}) to the Alfv\'{e}n Mach number (\ref{eq:dimless}) are therefore important here.

$M_{A2x}$ is plotted in terms of $M_{A1x}$ on the third row of Figure  \ref{fig:MHD}. The forbidden, non-evolutionary, zones have been colored on the plots. They feature the non-evolutionary transition just described, namely super-Alfv\'{e}nic $\rightarrow$ sub-Alfv\'{e}nic, together with the reverse transition sub-Alfv\'{e}nic $\rightarrow$ super-Alfv\'{e}nic. As a result, the MHD branches going through these regions are non-evolutionary, hence not physical solutions of the MHD problem.

The lowest row of figure \ref{fig:MHD} is the result of this evolutionary filter applied to the upper row. There is now but 1 solution for a given $M_{A1x}$, or none, as there are $M_{A1x}$-gaps where no solutions appear. Regardless of the initial conditions, the shock formed is never found with a $M_{A1x}$ inside such gaps like, for example, $M_{A1x}\in[1 , 1.4]$ for $\theta_1=0.5\pi/2$ (see Figure \ref{fig:MHD}, bottom right plot).

We shall  assume in the sequel that the evolutionarity criteria also applies in the collisionless case. This will have to be checked in future works.

\begin{figure}
\begin{center}
 \includegraphics[width=\textwidth]{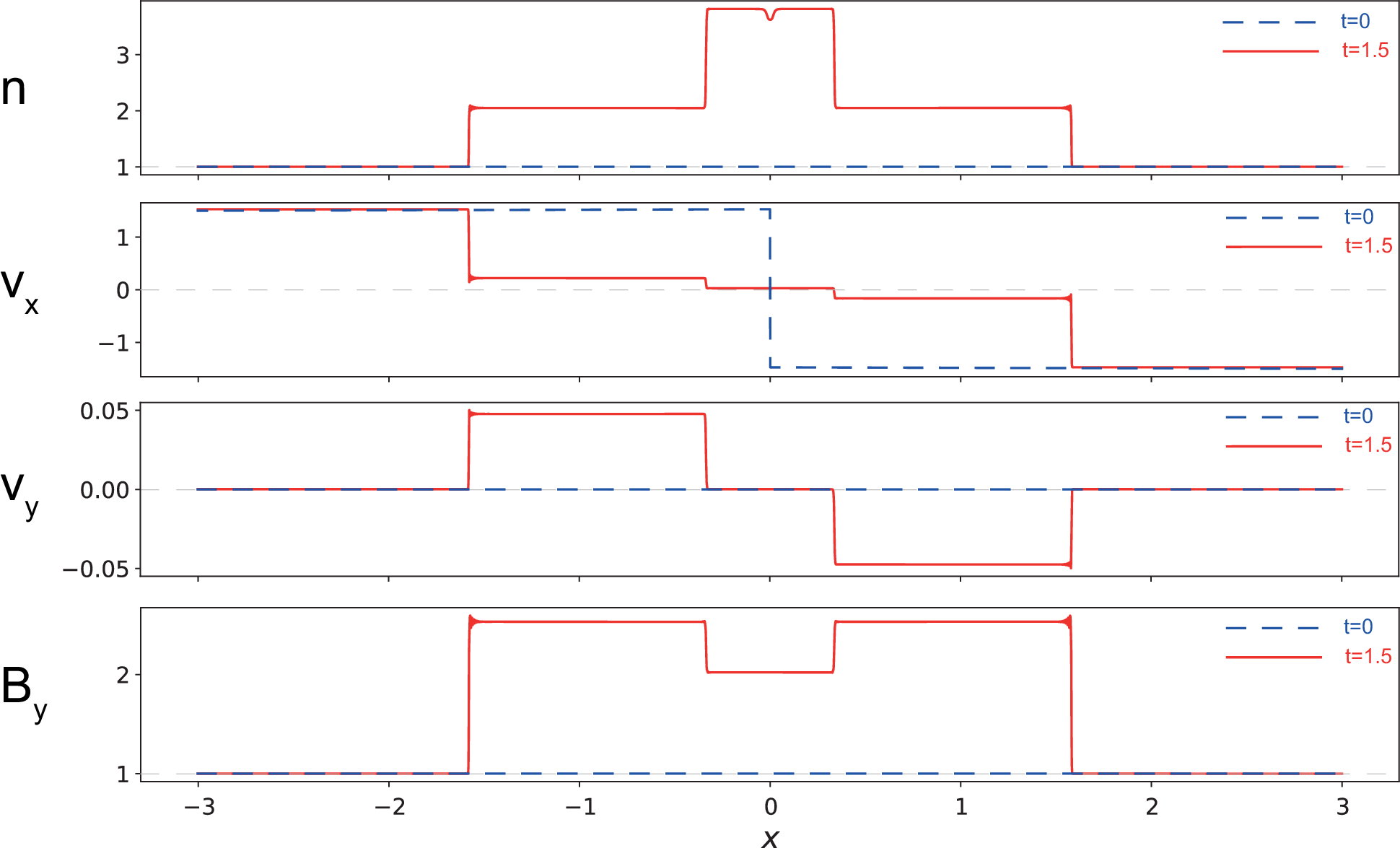}
\end{center}
\caption{MHD simulation in which the evolutionarity criterion results in the formation of 2 sub-shocks instead of one. The dashed blue lines show the initial fluid configuration at $t=0$, when two fluid columns collide at $x=0$. The solid red lines show the configuration at $t=1.5$. Each fluid column develops two shocks, a fast shock at $|x|=1.58$ and a slow shock at $|x|=0.34$. The little dip in density $n$ at $x=0$ for $t=1.5$ is a numerical artifact.}\label{fig:MHDSim}
\end{figure}

\subsection{MHD simulations}\label{eq:MHDsim}

In order to illustrate these theoretical results, we ran some MHD simulations with the code KORAL \citep{Olek2013,Olek2014}. KORAL is designed for multi-dimensional radiative MHD simulations in general relativistic spacetimes. However, if we turn off the radiation module as well as special and general relativity, the code reduces to a multi-dimenionsional non-relativistic MHD code. This version of the code was used here. Some of the results for $\theta_2$, $M_{A2x}$ and $r$ in terms of $M_{A1x}$ for upstream $\theta_1=0.5\pi/2$, are pictured on Figure \ref{fig:MHD} by the black circles. They perfectly line up with the theory and avoid the predicted gap in the solutions for $M_{A1x}\in[1 , 1.4]$.

Figure \ref{fig:MHDSim} shows the result of a simulation of a shock in a  non-evolutionary case. At $t=0$, two cold fluids of identical density and opposite velocities $\pm 1.5 \mathbf{x}$ collide (blue lines). The magnetic field has modulus unity and $\theta_1=0.5\pi/2$. At $t=1.5$, two shocks formed instead of one. The non-evolutionary case gives rise to 2 sub-shocks, as predicted in \cite{Kulsrud2005} for example.

\section{Instabilities involved and modified Alfv\'{e}n velocity}\label{sec:insta}
As previously stated, the starting point for our model are the MHD conservation equations with anisotropic temperatures. They have been derived in \cite{Hudson1970} and are reported in Appendix \ref{ap:MHDoblique} with the notations of \cite{BretJPP2022a,BretJPP2022b}.

For Stage 1, we solve them imposing relations (\ref{eq:ansatz}). This defines Stage 1. Then, mirror or firehose stability of Stage 1 has to be assessed. These instabilities are discussed here.

Now, evolutionarity involves the downstream Alfv\'{e}n velocity (projected onto the shock normal). As long as the plasma is isotropic, the Alfv\'{e}n velocity in given by Eq. (\ref{eq:Alfviso}). As a result, the downstream Alfv\'{e}n Mach number for the MHD switch-on shocks is exactly 1.

Yet, in an anisotropic plasma, it was found in \cite{AS67} that the Alfv\'{e}n velocity reads\footnote{This expression already includes the $\cos\theta$ factor of Eq. (\ref{eq:MAx}). There is no need to multiply it by an additional $\cos\theta$ when computing the Alfv\'{e}n Mach number (\ref{eq:MAx}).},
\begin{equation}\label{eq:AS67}
c_A = \pm v_A \sqrt{S_\perp - S_\parallel+1} ~ \cos\theta,
\end{equation}
where $v_A$ is still given by Eq. (\ref{eq:Alfviso}) and
\begin{equation}\label{eq:S_AS67}
S_{\perp,\parallel} = \frac{n k_B T_{\perp,\parallel}}{B^2/4\pi}.
\end{equation}

In Eq. (\ref{eq:AS67}), the $\pm$ sign refers to waves propagating along the field, or in the opposite direction. As shall be checked in the sequel (Sections  \ref{sec:theta0} \& \ref{sec:theta0PIC} and Figure \ref{fig:theta0}), with this correction to the Alfv\'{e}n velocity, the switch-on solutions of our model have $M_{A2x}=1$, exactly like in MHD.

In addition, the quantity below the square root can become negative. Such a situation indicates that the Alfv\'{e}n waves become unstable, which corresponds to the firehose instability. For this to happen, $S_\perp - S_\parallel+1 < 0$  is required, which can be cast under the form,
\begin{equation}\label{eq:fireAS67}
A \equiv \frac{T_\perp}{T_\parallel} < 1 - \frac{2}{\beta_\parallel},
\end{equation}
where\footnote{This parameter is just twice the $S_\parallel$ parameter defined by Eq. (\ref{eq:S_AS67}). We could use only $\beta_\parallel$. But for clarity, we present the result of \cite{AS67} with the notations of \cite{AS67}.},
\begin{equation}\label{eq:betapara}
\beta_\parallel = \frac{n k_B T_\parallel}{B^2/8\pi}.
\end{equation}

The criterion  (\ref{eq:fireAS67}) is therefore the one we shall use in the sequel, in order to preserve the inner coherence with the switch-on solutions of our model having $M_{A2x}=1$. This criterion  slightly differs, by the factor 2, from the one commonly used in the literature and in Paper I.

The Solar Wind data are a key test for the threshold of the firehose instability. They are indeed compatible with both criteria, namely with or without the factor 2 \citep{Hellinger2006,BalePRL2009,MarucaPRL2011}.

\begin{figure}
\begin{center}
 \includegraphics[width=\textwidth]{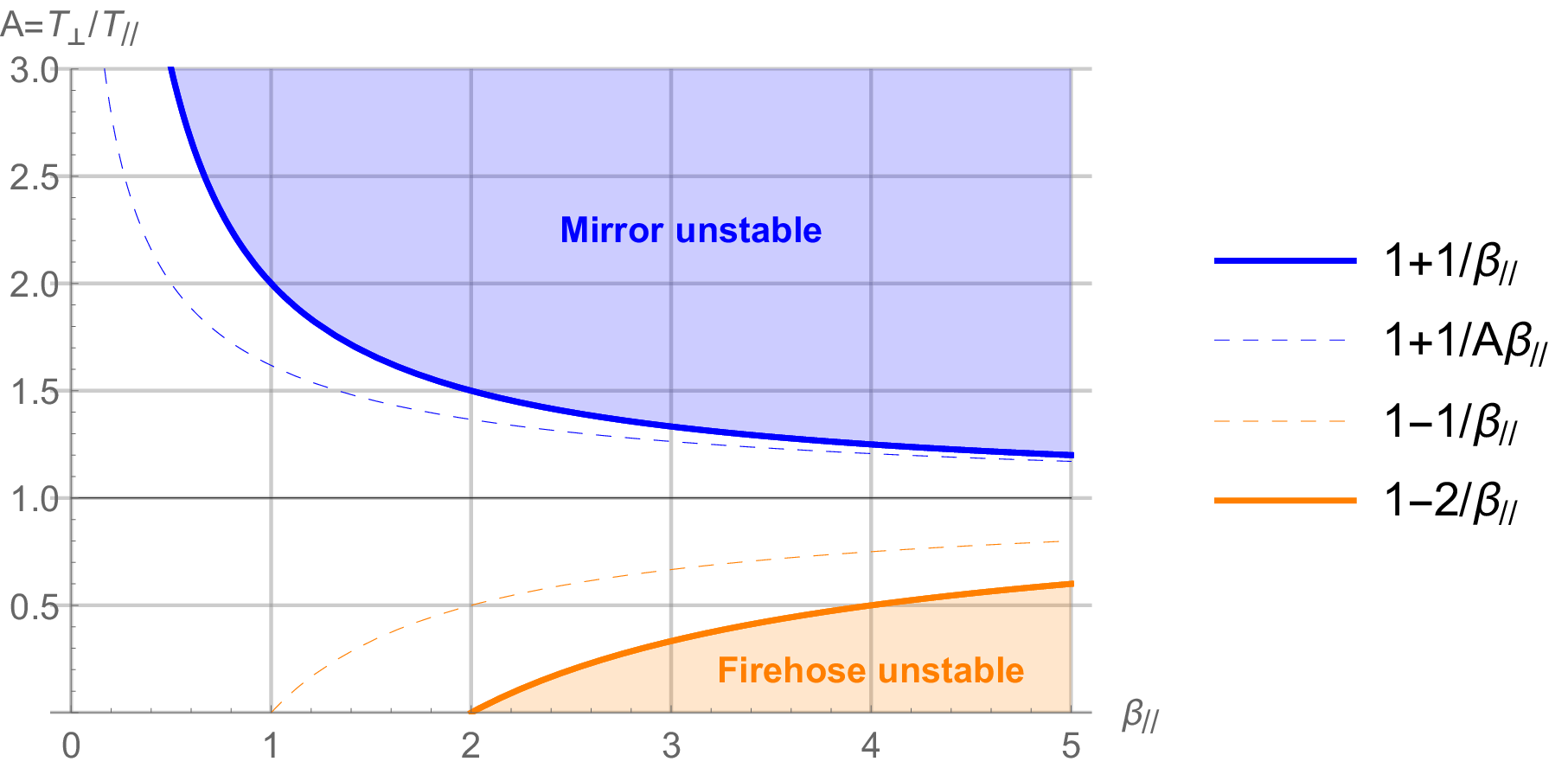}
\end{center}
\caption{Stability diagram. The solid lines picture the criterion  used here. The dashed lines pertain to other criteria discussed in Section \ref{sec:insta}. }\label{fig:stab}
\end{figure}

The other instability considered in relation to the stability of Stage 1, is the mirror instability. While the firehose instability occurs for too low an anisotropy $T_\perp/T_\parallel$, the mirror instability occurs for too high an anisotropy. The standard threshold given in the literature reads \citep{Hasegawa1975,Gary1993},
\begin{equation}\label{eq:mirror}
A > 1 + \frac{1}{\beta_\parallel}.
\end{equation}
Yet, a different criterion  is given in  \cite{AS67}, namely
\begin{equation}\label{eq:mirrorAS67}
A > 1 + \frac{1}{A ~\beta_\parallel}.
\end{equation}

However, while Stage 2-firehose remains analytically tractable using (\ref{eq:fireAS67}), Stage 2-mirror is not when using (\ref{eq:mirrorAS67}) instead of (\ref{eq:mirror}). Therefore, we adopt criterion  (\ref{eq:fireAS67}) for the firehose instability and keep (\ref{eq:mirror}) for mirror. Note in this respect that 1), the Solar Wind data are compatible with both criteria and 2), the inner coherence of the model does not impose a specific mirror criteria, as is the case for firehose.

Figure \ref{fig:stab} pictures the various criteria commented here for the mirror and the firehose instabilities. Even if the corrected criterion  for the mirror instability is functionally different than the one without the correction, the end result is qualitatively the same, and remains compatible with the Solar Wind data.

The two stability domains are disconnected, so that the plasma cannot be unstable to both at once. There is no possible competition between them.

\begin{figure}
\begin{center}
 \includegraphics[width=\textwidth]{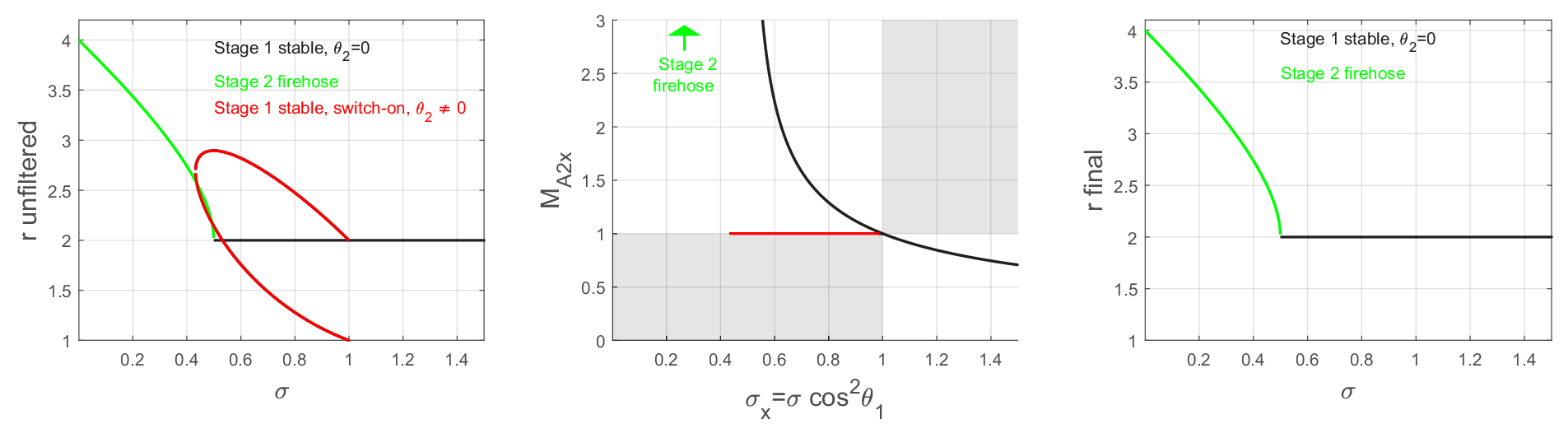}
\end{center}
\caption{\emph{Left}: All solutions offered by our model for $\theta_1=0$. \emph{Center}: All the solutions fulfill the evolutionarity criterion. Note that  this is different from the isotropic MHD problem, where some solutions for $\theta_1=0$ enter the shaded forbidden zone (Figure~\ref{fig:MHD}, row 3). Note also that, with definition (\ref{eq:AS67}) of the Alfv\'{e}n speed, $M_{A2x}=+\infty$ on the firehose threshold. This is why the green branch on the left is sent to $M_{A2x}=+\infty$ on the center plot. Notably, $M_{A2x}=1$ for the switch-on shocks of our model (red segment on center plot). For this analysis, the horizontal axis must be $\sigma_x=\sigma\cos^2\theta_1$, which for $\theta_1=0$ makes no difference. \emph{Right}: End result once the ``$\theta_2$ closest to $\theta_1$'' filter has been applied. This is the result found in \cite{BretJPP2018,Haggerty2022}.}\label{fig:theta0}
\end{figure}

\section{Branches selection}\label{sec:branch}\label{sec:selection}
Having specified the instabilities involved in the transition from Stage 1 to Stage 2, together with their respective thresholds, we can proceed to the filtering of the solutions our model offers. In this respect, it is instructive to  single out the case $\theta_1=0$.

\subsection{Case $\theta_1=0$}\label{sec:theta0}
Figure \ref{fig:theta0}-left shows the solutions offered by our model for $\theta_1=0$, without any filtering. It displays the various solutions with stable Stage 1, plus a branch, in green, corresponding to Stage 2-firehose because Stage 1 is firehose unstable for $\sigma \in [0 , 0.5]$. For $\sigma \in [0.5,1]$, there are up to 3 possible solutions: one has $r=2$ and $\theta_2 = 0$, and the other two pertain to the switch-on solutions in red, with $\theta_2 \neq 0$.

Could the evolutionarity criterion  help filtering them (the answer is `no') ? Figure \ref{fig:theta0}-center answers the question. Here is plotted the downstream Mach number $M_{A2x}$ of each solution, as defined by Eq. (\ref{eq:MAx}), where the Alfv\'{e}n speed has been corrected according to Eq. (\ref{eq:AS67}). Our switch-on solutions have exactly $M_{A2x} = 1$, while the others also satisfy the evolutionary criterion\footnote{Figure 5(b) of \cite{BretJPP2022a} shows the switch-on solutions of our model with $M_{A2x} \neq 1$. This is because correction (\ref{eq:AS67}) to the Alfv\'{e}n speed was not considered.}. In line with the adequate definition (\ref{eq:MAx}) of the Alfv\'{e}n Mach number for evolutionarity analysis, we use on the horizontal axis the parameter
\begin{equation}\label{eq:sigmax}
\sigma_x \equiv \sigma \cos^2\theta_1,
\end{equation}
which, for the present case $\theta_1=0$, makes no difference.

Note that with definition (\ref{eq:AS67}) of the Alfv\'{e}n speed, $M_{A2x}=+\infty$ on the firehose threshold. This is the reason why the green branch on the left is sent to $M_{A2x}=+\infty$ on the center plot. As a consequence, each  time the system eventually settles in Stage 2-firehose with $\sigma_x < 1$ (i.e, $M_{A1x}>1$), it is evolutionary.

As a conclusion, in the interval $\sigma \in [0.5,1]$, there are three solutions for Stage 1, and all three satisfy the evolutionarity criterion. The evolutionarity criterion by itself is therefore \emph{not} sufficient to trim the number of solutions down to 1, or even 0.

\begin{figure}
\begin{center}
 \includegraphics[width=0.45\textwidth]{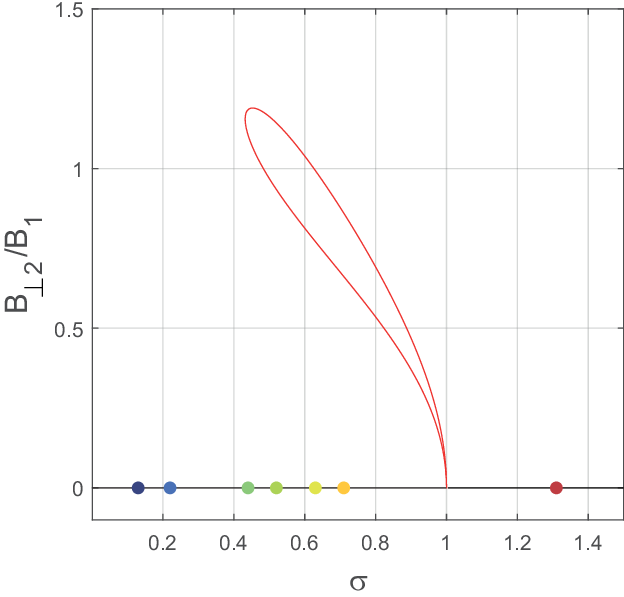} \includegraphics[width=0.45\textwidth]{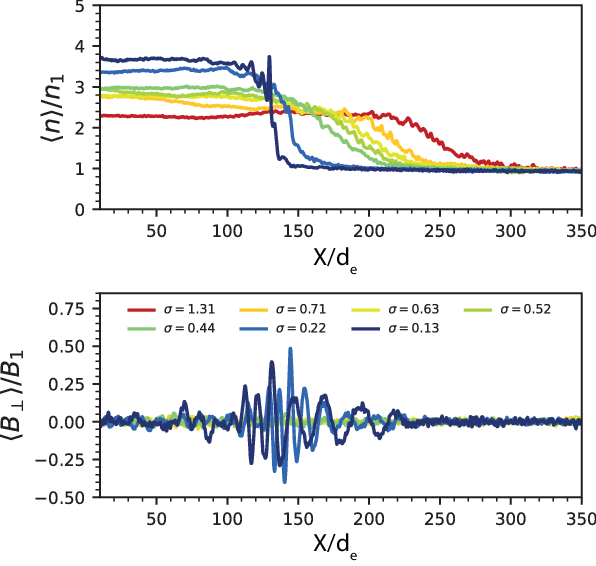}
\end{center}
\caption{\emph{Left}: expected value of $B_{\perp 2}/B_1$, in red for the switch-on solutions with $\theta_2\neq 0$, and in black for the solution with $\theta_2=\theta_1=0$. The circles show the values of $\sigma$ which have been simulated. \emph{Right}: results of the PIC simulations, with the shock density profile on top, and the ratio $B_{\perp 2}/B_1$ below. The $\sigma$ value for a curve is given by the circle of the same color on the left plot. The same results are obtained for $\theta_1=2^\circ$. The flow is along the $x$ axis and $d_e=c/\omega_p$ is the electron inertial length.}\label{fig:sigmatested}
\end{figure}

\subsection{PIC simulations for $\theta_1=0$}\label{sec:theta0PIC}
In order to check which solution the system chooses, we ran a series a PIC simulations for various $\sigma$'s, performed with the fully kinetic 3D PIC code, TRISTAN-MP \citep{Buneman1993,Spitkovsky2005}\footnote{Simulation parameters identical to those of \cite{Haggerty2022}.}. The surest way to tell whether the shock is of the switch-on type or not, is to plot the perpendicular component $B_{\perp 2}$ of the field in the downstream. According to Eq. (\ref{eq:conser2}), with $\theta_1=0$ it simply reads
\begin{equation}\label{eq:B2perp}
B_{\perp 2} = B_1 \tan \theta_2.
\end{equation}

Figure \ref{fig:sigmatested}-left shows the expected value of $B_{\perp 2}/B_1$, in red for the switch-on solutions with $\theta_2\neq 0$, and in black for the solution with $\theta_2=0$. The circles show the values of $\sigma$ which have been simulated. Figure \ref{fig:sigmatested}-right shows the results of the PIC simulations, with the shock density profile on top, and the ratio $B_{\perp 2}/B_1$ below. Besides some perturbations around the shock front at low $\sigma$, $B_{\perp 2}/B_1$ does not depart from 0 in the downstream for any $\sigma$. These perturbations are due to particle acceleration and back-reaction in the precursor, and from the instabilities triggered by the interaction of the fast upstream flow with the front \citep{Sironi2013ApJ,BrownJPP2023}.

Simply put, PIC simulations consistently discard the switch-on solutions. The same results are obtained for $\theta_1=2^\circ$, so that we are not witnessing a singular behavior fruit of a perfect, hence unrealistic, symmetry. A similar pattern was observed in the relativistic regime in \cite{BretJPP2017}.

The situation for $\theta_1 \sim 0$ is here markedly different from the MHD case. In MHD, where isotropy is imposed, the evolutionarity criterion imposes switch-on shocks within some $\sigma$-range, as explained in Section \ref{sec:MHD}. In our model, where an anisotropy is driven by the field, the evolutionarity criterion does \emph{not} impose switch-on solutions, while PIC simulations consistently choose the non switch-on ones.

Arguably this explains why so few detections of switch-on shocks have been made in the Solar System. Indeed, among the thousands of shocks observed (\cite{Farris1994,Russell1995}, \cite{Balogh2013} \S 2.3.6.) only one ``possible'' detection of an interplanetary switch-on shock was reported in \cite{Feng2009}. Also, \cite{Russell1995} reported the detection of only one switch-on shock among the International Sun-Earth Explorer data.\footnote{Switch-on shocks have been produced in the laboratory  \citep{Craig1973}, but within a collisional environment, where MHD rules. We here deal with collision\emph{less} plasmas.}

Still, what about these exceptions, since our collisionless scenario should rule them out? Several explanations are possible,
\begin{itemize}
  \item The detections were faulty. Hence the adjective ``possible'' associated with one of them.
  \item We only solve our model for a cold upstream, namely $T_1=0$. Maybe a finite $T_1$ would have our model allowing for some switch-on shocks.
  \item The \emph{ansatz} (\ref{eq:ansatz}) is not accurate enough, and a better version would allow for some switch-on shocks.
\end{itemize}

At any rate, PIC simulations and observations tell switch-on shocks in collisionless plasma are rare. We therefore propose the following criterion allowing to choose between various solutions:\emph{ the system chooses the solution with $\theta_2$ closest to $\theta_1$.}

Note that this ``$\theta_2$ closest to $\theta_1$'' criterion stems from our PIC simulations, as well as others of parallel shocks in pair \citep{BretJPP2017,Haggerty2022} or electron/ion  \citep{niemiec2012,Zekovic2019} plasmas, where the same, non switch-on branch, is consistently chosen. This is why we used the verb ``propose''. Its robustness on longer time scales, or other shock geometries, is beyond the scope of this work and should be assessed in further works.

Figure \ref{fig:theta0}-right eventually shows the end result once the ``$\theta_2$ closest to $\theta_1$'' filter has been applied to the $\theta_1=0$ case. There is now but one solution for any $\sigma$, which indeed is the one found in \cite{BretJPP2018,Haggerty2022}.

\begin{figure}
\begin{center}
 \includegraphics[width=\textwidth]{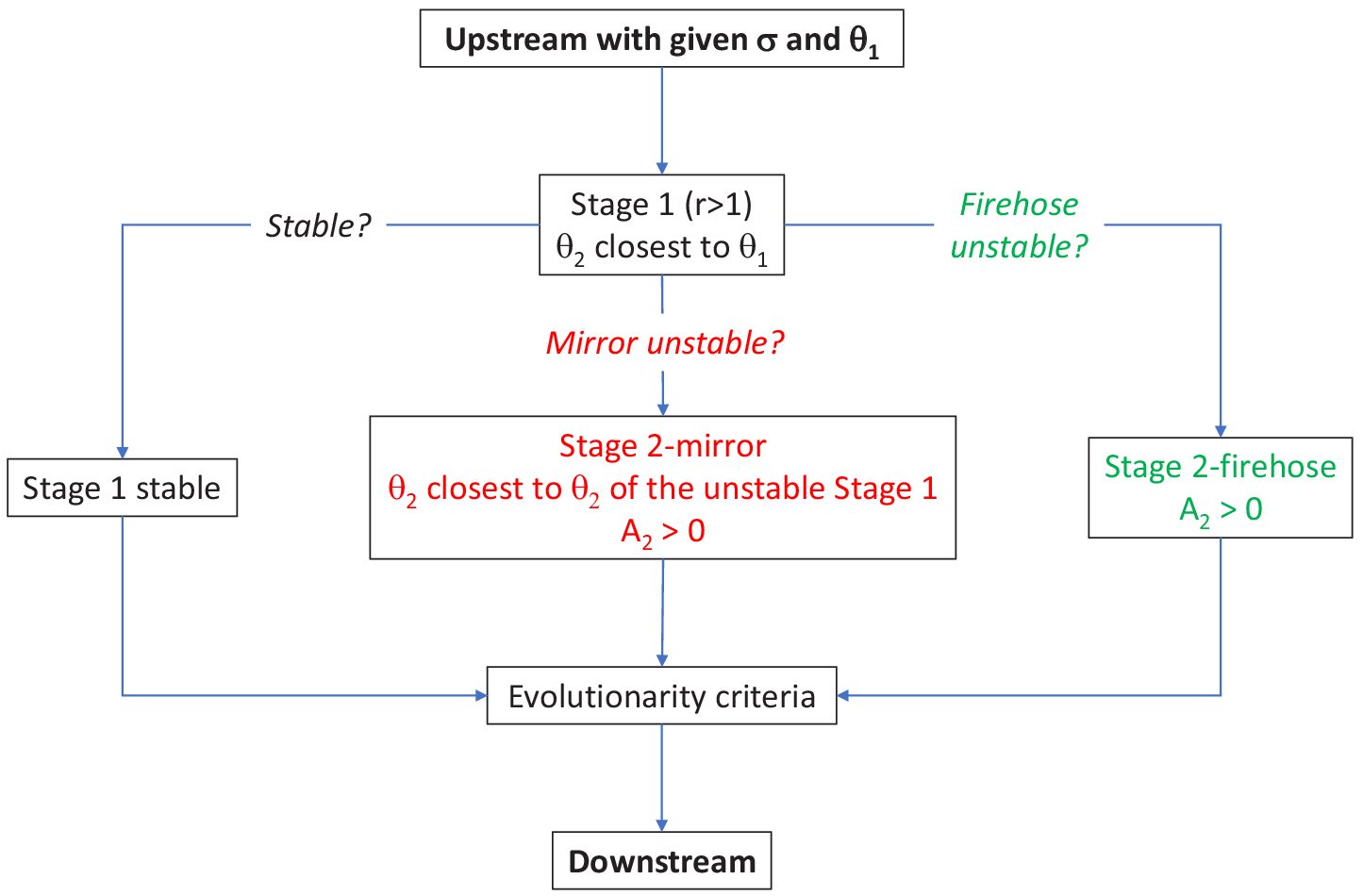}
\end{center}
\caption{Flowchart of the resolution of our model. In case Stage 1 is mirror unstable for a pair $(\sigma,\theta_1)$, there can be a degeneracy in the choice of Stage 2-mirror for the same pair $(\sigma,\theta_1)$. In such a case, we choose the Stage 2-mirror with the $\theta_2$ closest to the $\theta_2$ of the Stage 1 it comes from, as was already done in \cite{BretJPP2022b}. Such a situation never happens with Stage 2-firehose.}\label{fig:flowchart}
\end{figure}

\subsection{General algorithm for branches selection}
We may now lay out the general algorithm for branches selection. The criteria used to eliminate some are, applied in this order,
\begin{enumerate}
  \item Exclude Stage 1 solutions with $r<1$. Then select the one with the $\theta_2$ closest to $\theta_1$.
  \item In case Stage 1 is unstable, is the anisotropy $A_2$ of a given Stage 2 solution, negative? This was already implemented in Paper I and allows to eliminate some Stage 2 -firehose and -mirror solutions. This never happens with Stage 1 since it has, by design from Eqs. (\ref{eq:ansatz}), $A_2 = \frac{1}{2}\tan^2\theta_2$.
  \item Does the resulting solution fulfill the evolutionarity criterion ?
\end{enumerate}

The evolutionarity criterion  is applied last because it operates on time scales related to the propagation of the shock\footnote{See Section 6.3 in \cite{FalleJPP2001}.}, whereas the other criteria operate on much shorter time scales, related to plasma instabilities.

The algorithm is eventually represented by the flowchart on Figure \ref{fig:flowchart}. In case Stage 1 is mirror unstable for a given pair $(\sigma,\theta_1)$, there can be a degeneracy in the choice of the Stage 2-mirror states for the same pair $(\sigma,\theta_1)$. In such a case, we choose the Stage 2-mirror with the $\theta_2$ closest to the $\theta_2$ of the Stage 1 it comes from, as was already done in \cite{BretJPP2022b}. Such a situation never happens with Stage 2-firehose.

According to the flowchart on Figure \ref{fig:flowchart}, there is necessarily only 1 solution left before applying the evolutionarity criterion.  There is therefore no need to apply the ``$\theta_2$ closest to $\theta_1$'' filter in (\emph{c}), since only 1 branch can make it to this stage. Yet, this does not mean the evolutionarity filter is useless, as it can simply forbid the existence of a solution in some $\sigma$ range, as is the case in MHD (see bottom row of Figure \ref{fig:MHD}).

\begin{figure}
\begin{center}
\includegraphics[width=\textwidth]{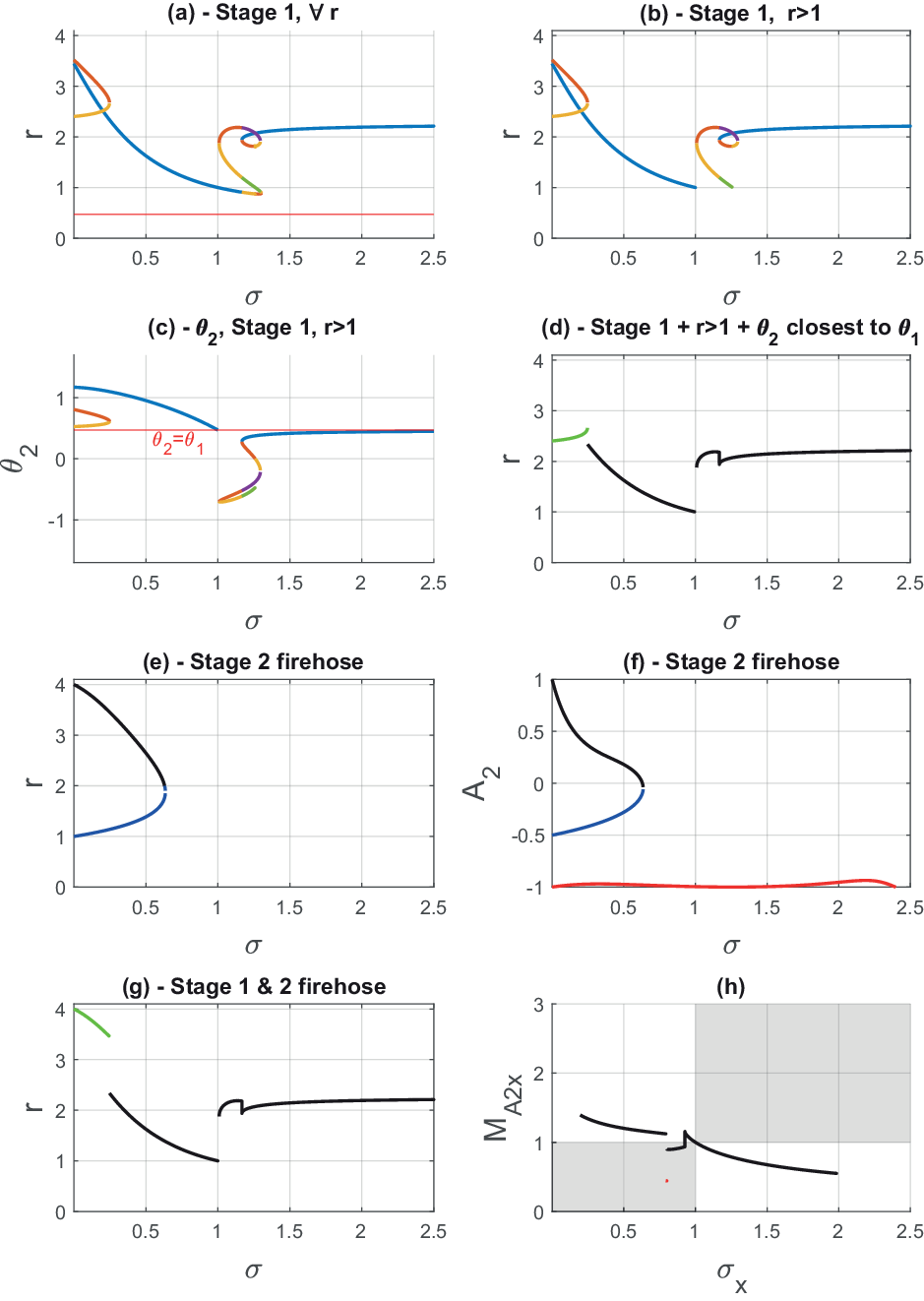}
\end{center}
\caption{Detailed progression of the application of the filtering process described by the flow chart on Figure \ref{fig:flowchart}, for $\theta_1=0.3\pi/2$. On Figure (d), the green color indicates that these Stage 1 solutions  are firehose unstable. Then on Figure (g), the green color shows the density jump of Stage 2-firehose. The red branch on Figure (f) with $A_2 \sim -1$ discards Stage 2 firehose solutions with $r >4$, not shown on Figure (e).}\label{fig:resultoblique}
\end{figure}

\section{Results}
While the fruit of our algorithm has already been explained for $\theta_1=0$, it is interesting to detail how it unfolds for an oblique field, like for example $\theta_1 = 0.3\pi/2$.

Figure \ref{fig:resultoblique}(a) shows all possible density jumps for Stage 1. Notably, $r < 1$ for $\sigma \in [1, 1.3]$. The corresponding branches are then eliminated on Figure \ref{fig:resultoblique}(b).

For these Stage 1 solutions with $r > 1$, Figure \ref{fig:resultoblique}(c) then shows the corresponding values of $\theta_2$. The horizontal red line is at $\theta_2=\theta_1$. The result of the ``$\theta_2$ closest to $\theta_1$'' selection rule is displayed on Figure \ref{fig:resultoblique}(d).

On Figure \ref{fig:resultoblique}(d), the green color indicates that some Stage 1 solutions at low $\sigma$ are firehose unstable. We therefore need to examine Stage 2-firehose solutions. All of them are displayed on Figure \ref{fig:resultoblique}(e). Yet, Figure \ref{fig:resultoblique}(f) shows that the lower branch has $A_2 < 0$ and needs to be eliminated. As a consequence, the only physical Stage 2-firehose solution available when Stage 1 is firehose unstable, is the upper branch.

Replacing then the firehose unstable Stage 1 solutions of Figure \ref{fig:resultoblique}(d), by the corresponding Stage 2-firehose solutions, gives Figure \ref{fig:resultoblique}(g).

We finally need to apply the evolutionarity test to the solutions of Figure \ref{fig:resultoblique}(g). This is done on Figure \ref{fig:resultoblique}(h) where the downstream Alfv\'{e}n Mach number $M_{A2x}$ has been computed for the solution plotted on Figure \ref{fig:resultoblique}(g), with the forbidden zones shaded. Note that for such an analysis, the horizontal scale has to be $\sigma_x=\sigma\cos^2\theta_1$. Then and only then has the evolutionary analysis some branches passing exactly through the point $(1,1)$, like on Figure \ref{fig:resultoblique}(h).

\begin{figure}
\begin{center}
\includegraphics[width=\textwidth]{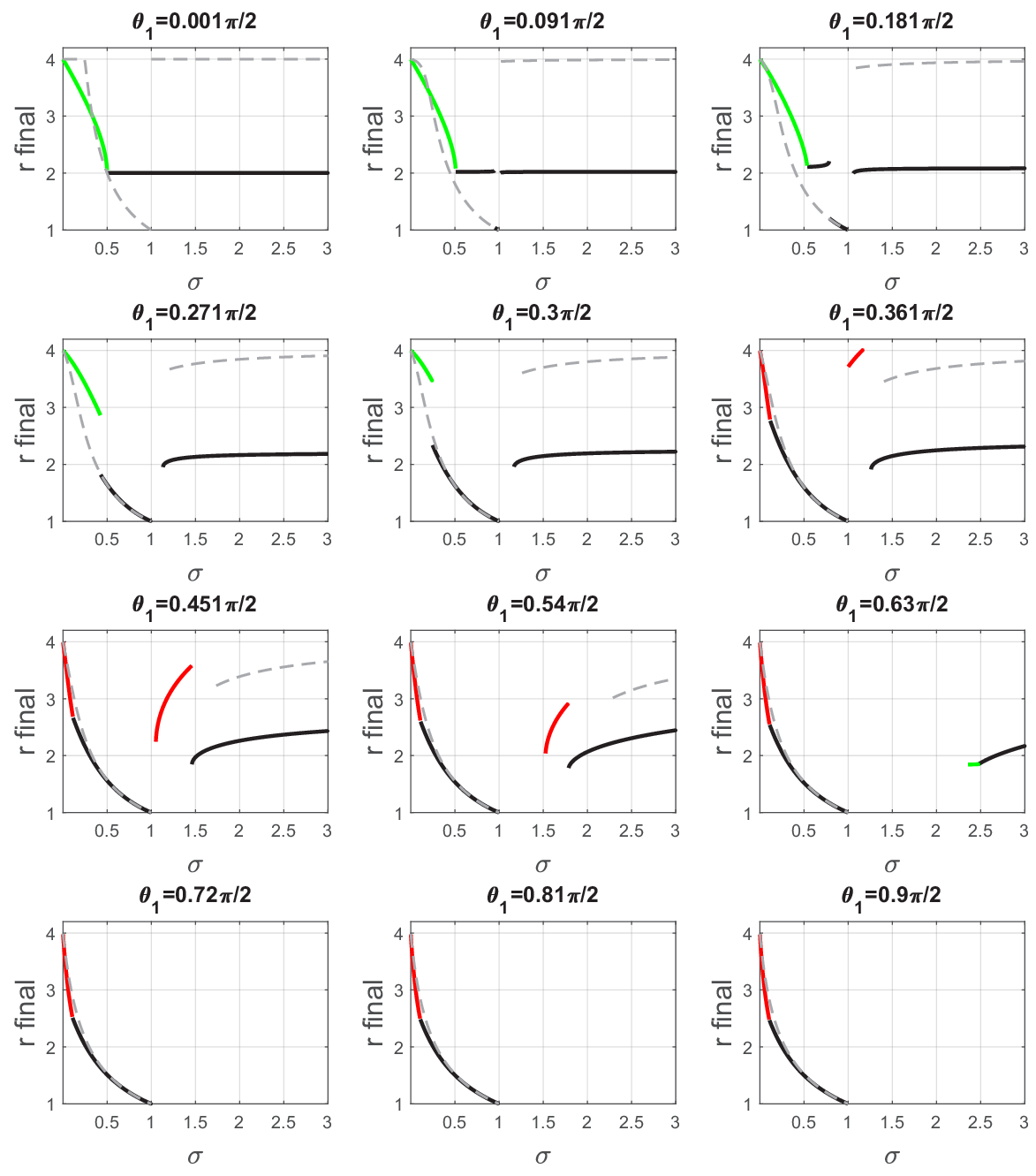}
\end{center}
\caption{Result of the filtering process elaborated in this work for various angles $\theta_1 \in [0, \pi/2]$. All of them but $\theta_1=0.3\pi/2$, are the ones considered in Figure 8 of \cite{BretJPP2022b}. The gray dashed lines picture the MHD result, evolutionary-filtered. The green solutions pertain to \textcolor[rgb]{0.00,1.00,0.00}{Stage 2-firehose}, and the red ones to \textcolor[rgb]{1.00,0.00,0.00}{Stage 2-mirror}.}\label{fig:rFinal}
\end{figure}

The Stage 2-firehose branch visible in green on Figure \ref{fig:resultoblique}(g) at small $\sigma$ passes the evolutionarity test since it has $\sigma_x < 1$ and $M_{A2x} = +\infty$. The analysis shows that some stable Stage 1 solutions do not pass the evolutionarity test. As a result, Figure \ref{fig:rFinal} for $\theta_1=0.3\pi/2$ displays a gap without solution for $\sigma \in [1,1.17]$, that does not show on Figure \ref{fig:resultoblique}(g).

The end result of such a filtering process is eventually shown on Figure \ref{fig:rFinal} for several angles $\theta_1$ between 0 and $\pi/2$. For comparison, all of them but $\theta_1=0.3\pi/2$, are the ones considered in Figure 8 of \cite{BretJPP2022b}. The gray dashed line pictures the MHD result, evolutionary-filtered.

Figure \ref{fig:rFinal} eventually interpolates between the parallel case treated in \cite{BretJPP2018}, and the perpendicular one treated in \cite{BretPoP2019}. While Section \ref{sec:selection} showed that some filtering is needed to get to the end result for the parallel case, no filtering at all is required for the perpendicular case. Figure \ref{fig:rFinal} for $\theta_1=0.9\pi/2$ gives the result already found in \cite{BretPoP2019}, without any filtering but the $A_2 < 0$ for the mirror solutions. The main reason for this is that for the perpendicular case, there is but one Stage 1 solution, which,  when   stable, fulfills the evolutionarity criterion.

\section{Conclusion}
Applying the MHD formalism to analyse collisionless shocks may be problematic in the presence of an ambient magnetic field, capable of stabilizing pressure anisotropies. In a series of recent papers, we developed a method allowing to determine the downstream anisotropy of a collisionless shock  \citep{BretJPP2018,BretPoP2019,BretJPP2022a,BretJPP2022b,BretMNRAS2023a,BretMNRAS2023b}. The anisotropy can then be included in the MHD conservation equations for anisotropic pressures, and the modified density jump derived. The case of a parallel shock has been successfully tested by PIC simulations in \cite{Haggerty2022}, confirming that for a strong enough field, the density jump can go from 4, the expected MHD value, to only 2, the anisotropy corrected one.

Once the density jump is found, all the other jumps like pressure, temperature, entropy, etc., can be straightforwardly derived \citep{BretPoP2021}.

As can be seen on Figure \ref{fig:rFinal}, our results differ from the isotropic MHD ones in 2 ways,
\begin{itemize}
  \item The ranges of $\sigma$ without solutions differ.
  \item For $\sigma$ values with a solution, our density jump is usually, though not always, lower than the MHD one.
\end{itemize}

Overall, our results and the MHD ones bear a ``family resemblance'', the largest discrepancy being found for the parallel case $\theta_1=0$. The predicted large reduction of the density jump could have important consequences for particle acceleration, since their index scales like $(r-1)^{-1}$ \citep{Blandford78,Axford1977,Bell1978a,Bell1978b,Haggerty2020a,Haggerty2020b}.

Besides the strong sonic shock assumption of the present work, namely $P_1=0$, its main limitation may lie in the composition of the plasma, here a pair plasma. As stated in the introduction, such an assumption allows to consider only one parallel and one perpendicular temperature. Yet, our formalism being eventually macroscopic, we expect our conclusions to hold for electron/ion plasmas as well. Such a conjecture is currently being tested through PIC simulations of such plasmas \citep{Shalaby2022}.

\section*{Acknowledgments}
Thanks are due to  Alexander Velikovich and Federico Fraschetti for valuable inputs.

\section*{Funding}
A.B. acknowledges support by the Ministerio de Econom\'{i}a  y Competitividad of Spain (Grant No. PID2021-125550OB-I00).
R.N. acknowledges support from the NSF Grant No. AST-1816420. C.C.H. was partially supported by NSF FDSS grant AGS-1936393 as well as NASA grants 80NSSC20K1273 and 80NSSC23K0099.

Simulations were performed on computational resources provided by TACC’s Stampede2 and Purdue's ANVIL through ACCESS (formally XSEDE) allocation TG-AST180008. Some of the work was supported by the Geospace Environment Modeling Focus Group ``Particle Heating and Thermalization in Collisionless Shocks in the Magnetospheric MultiScale Mission (MMS) Era''.

R.N. and A.B. thank the Black Hole Initiative at Harvard University for support. The BHI is funded by grants from the John Templeton Foundation and the Gordon and Betty Moore Foundation.

\section*{Declaration of Interests}
The authors report no conflict of interest.

\appendix
\section{Isotropic MHD conservation equations for an oblique shock}\label{ap:MHD}
The MHD conservation equations for an oblique shock and a fluid of adiabatic index $\gamma=5/3$ read \citep{Kulsrud2005},
\begin{eqnarray}
n_2 v_2 \cos  \xi _2  &=& n_1 v_1,  \label{eq:consmhd1} \\
B_2 \cos \theta_2 &=& B_1 \cos \theta_1  ,  \label{eq:consmhd2}\\
B_2 v_2 \sin \theta_2 \cos \xi_2 - B_2 v_2 \cos \theta_2 \sin \xi_2 &=&  B_1 v_1 \sin \theta_1 ,  \label{eq:consmhd3}\\
\frac{B_2^2 \sin ^2\theta_2}{8 \pi }+n_2 k_B T_2 + m n_2 v_2^2 \cos ^2\xi_2
                                                  &=& \frac{B_1^2 \sin ^2\theta_1}{8 \pi }+n_2 k_B T_1 + m n_1 v_1^2 , \label{eq:consmhd4}\\
- \frac{B_2^2}{4 \pi }\sin \theta_2 \cos \theta_2 + m n_2 v_2^2 \sin \xi_2 \cos \xi_2 &=& - \frac{B_1^2 \sin \theta_1 \cos \theta_1}{4 \pi } , \label{eq:consmhd5}\\
\mathcal{A} v_2 \sin \xi_2 +   \mathcal{C} &=& m n_1 v_1 \left(\frac{5 k_B T_1 }{2 m}+\frac{B_1^2 \sin ^2\theta_1}{4 \pi  m n_1}+\frac{v_1^2}{2}\right), \label{eq:consmhd6}
\end{eqnarray}
where  $m$ is the mass of the particles, $k_B$ the Boltzmann constant, and
\begin{eqnarray}
\mathcal{A} &=&  - \frac{B_2^2}{4 \pi }\sin \theta_2 \cos \theta_2, \nonumber\\
\mathcal{C} &=& m n_2 v_2 \cos \xi_2 \left( 5\frac{k_BT_2}{2 m}  +\frac{B_2^2 \sin ^2\theta_2}{4 \pi  m n_2}+\frac{v_2^2}{2}\right).\nonumber
\end{eqnarray}

After some algebra, $\overline{T}_2 \equiv \tan\theta_2$ is found solution of\footnote{This quantity is written here with a bar to avoid confusion with the downstream temperature $T_2$ in Eqs. (\ref{eq:consmhd4},\ref{eq:consmhd6}). Such a confusion is excluded in the rest of the paper since the downstream temperature splits into 2 different quantities, namely $T_{\parallel 2}$ and $T_{\perp 2}$.}

\begin{equation}\label{eq:T2MHD}
\sum_{k=0}^4 a_k \overline{T}_2^k = 0,
\end{equation}
with,
\begin{eqnarray}
  a_0 &=& 2 M_{A1}^2 \left(\sin 2\theta_1-2 M_{A1}^2 \tan \theta_1\right)^2, \nonumber\\
  a_1 &=& -\frac{1}{8} M_{A1}^2 \tan \theta_1 \left[-4 M_{A1}^2 \left(31 \cos 2\theta_1+21\right)+80 M_{A1}^4+4 \cos ^2\theta_1 \left(21 \cos 2\theta_1+11\right)\right], \nonumber\\
  a_2 &=& \frac{1}{2} M_{A1}^2 \cos ^2\theta_1 \left(15 \cos 2\theta_1+1\right)-M_{A1}^4 \left(7 \cos 2\theta_1+3\right)+2 M_{A1}^6, \nonumber\\
  a_3 &=& -\frac{1}{4} \sin 2\theta_1 M_{A1}^2 \left(-2 M_{A1}^2+\cos 2\theta_1+1\right), \nonumber\\
  a_4 &=& 3 \cos ^4\theta_1 M_{A1}^2 .
\end{eqnarray}

\section{Anisotropic MHD  conservation equations for an oblique shock}\label{ap:MHDoblique}
The conservation equations for anisotropic temperatures were derived in \cite{Hudson1970,Erkaev2000}. They have been re-derived in \cite{BretJPP2022a} with the present notations. They are formally valid even for anisotropic upstream temperatures, with $T_{\parallel 1} \neq T_{\perp 1}$. Writing them for $T_{\parallel 1}=T_{\perp 1}\equiv T_1$, they read,
\begin{eqnarray}
n_2 v_2 \cos  \xi _2  &=& n_1 v_1,  \label{eq:conser1}  \\
B_2 \cos \theta_2 &=& B_1 \cos \theta_1  ,  \label{eq:conser2} \\
B_2 v_2 \sin \theta_2 \cos \xi_2 - B_2 v_2 \cos \theta_2 \sin \xi_2 &=&  B_1 v_1 \sin \theta_1 , \label{eq:conser3} \\
\frac{B_2^2 \sin ^2\theta_2}{8 \pi }+n_2 k_B ( T_{\parallel 2} \cos ^2\theta_2 + T_{\perp 2} \sin ^2\theta_2) + m n_2 v_2^2 \cos ^2\xi_2
                                                  &=& \frac{B_1^2 \sin ^2\theta_1}{8 \pi }+n_2 k_B T_1 + m n_1 v_1^2 , \label{eq:conser4}\\
\mathcal{A} + m n_2 v_2^2 \sin \xi_2 \cos \xi_2 &=& - \frac{B_1^2 \sin \theta_1 \cos \theta_1}{4 \pi } , \label{eq:conser5}\\
\mathcal{A} v_2 \sin \xi_2 + \mathcal{B} + \mathcal{C} &=& m n_1 v_1 \left(\frac{5 k_B T_1 }{2 m}+\frac{B_1^2 \sin ^2\theta_1}{4 \pi  m n_1}+\frac{v_1^2}{2}\right), \label{eq:conser6}
\end{eqnarray}
where
\begin{eqnarray}
\mathcal{A} &=& \sin \theta_2 \cos \theta_2 n_2 k_B \left(T_{\parallel 2}-T_{\perp 2}\right) - \frac{B_2^2}{4 \pi }\sin \theta_2 \cos \theta_2, \nonumber\\
\mathcal{B} &=& v_2 \cos ^2\theta_2 \cos \xi_2 n_2 k_B ( T_{\parallel 2} -   T_{\perp 2} ),  \nonumber\\
\mathcal{C} &=& m n_2 v_2 \cos \xi_2 \left( \frac{k_B}{2 m}(T_{\parallel 2} + 4 T_{\perp 2} )  +\frac{B_2^2 \sin ^2\theta_2}{4 \pi  m n_2}+\frac{v_2^2}{2}\right).\nonumber
\end{eqnarray}
It can be checked that setting $T_{\parallel 2} =   T_{\perp 2} = T_2$ gives back the MHD Eqs. (\ref{eq:consmhd1}-\ref{eq:consmhd6}).

\section{Main quantities for Stages 1 \& 2}\label{ap:quantities}
Implementing the algorithm described in Figure \ref{fig:flowchart} requires computing the density ratio $r$, the anisotropy $A_2$ and the downstream Alfv\'{e}n Mach number $M_{A2x}$, for Stage 1, Stage 2-firehose and Stage 2-mirror. The results are presented below.

\subsection{Stage 1}
Solving the system of equations (\ref{eq:conser1}-\ref{eq:conser6}) with prescription (\ref{eq:ansatz}) allows to  derive a polynomial for the quantify $T_2=\tan\theta_2$ defined in (\ref{eq:T_2}). It has been derived in Paper I\footnote{See Eqs. (4.2-4.5) in Paper I.}. Using now the expression (\ref{eq:AS67}) for the Alfv\'{e}n velocity, the Alfv\'{e}n Mach number for Stage 1 reads,
\begin{equation}\label{eq:MA2S1}
M_{A2x}^2 = \frac{4 \left(T_2^4+2\right) \sec^2\theta_1 M_{A1}^2}{\left(T_2^2-2\right) \sec^2\theta_1 \left(4 M_{A1}^2 (r-1)-r \cos 2 \theta_1+r\right)+2 r \left(T_2^4+2 T_2^2+4\right)}.
\end{equation}
From Eqs. (\ref{eq:ansatz}), the anisotropy of Stage 1 is simply,
\begin{equation}
A_2 = \frac{1}{2}\tan^2\theta_2.
\end{equation}
The density ratio $r$ is the same as in Paper I, namely,
\begin{equation}\label{eq:rS1}
r=\frac{4 \mathcal{M}_{A1}^2 T_2^3 (1+T_2^2)}{\sum_{k=0}^5b_k T2^k},
\end{equation}
where,
\begin{eqnarray}
  b_0 &=& 8 \mathcal{M}_{A1}^2 \tan \theta_1-4 \sin 2 \theta_1, \nonumber\\
  b_1 &=& -8 \mathcal{M}_{A1}^2+6 \cos 2 \theta_1+2, \nonumber\\
  b_2 &=& 0, \nonumber\\
  b_3 &=& 4 \mathcal{M}_{A1}^2+\cos 2 \theta_1+3, \nonumber\\
  b_4 &=& 4 \mathcal{M}_{A1}^2 \tan \theta_1-2 \sin 2 \theta_1, \nonumber\\
  b_5 &=&  2 \cos ^2\theta_1.
\end{eqnarray}

\subsection{Stage 2 firehose}
If Stage 1 has $A_2 < 1-2/\beta_{\parallel 2}$, then Stage 2-firehose is the end state. Since the stability criterion differs from that used in Paper I by the factor 2, the properties of Stage 2 firehose change with respect to Paper I. A polynomial equation can still be derived for $T_2=\tan\theta_2$ as,
\begin{equation}\label{eq:S2f}
\sum_{k=0}^3 a_k T_2^k = 0,
\end{equation}
with,
\begin{eqnarray}
  a_0 &=& -8 \left(\sin 2 \theta_1-2 M_{A1}^2 \tan \theta_1\right)^2 \\
  a_1 &=& \left(-2 M_{A1}^2+\cos 2 \theta_1+1\right) \left(-20 M_{A1}^2+9 \cos 2 \theta_1-1\right) \tan \theta_1  \\
  a_2 &=&  8 \cos 2 \theta_1 M_{A1}^2-8 \left(M_{A1}^4+M_{A1}^2\right)-\cos 4 \theta_1+1  \\
  a_3 &=&   \left(2 M_{A1}^2-\cos 2 \theta_1-1\right) \sin 2 \theta_1 .
\end{eqnarray}
The density jump is then given by
\begin{equation}\label{eq:rS2m}
r=\frac{M_{A1}^2 T_2}{M_{A1}^2 \tan \theta_1-\sin \theta_1 \cos \theta_1}.
\end{equation}
As for the Alfv\'{e}n Mach number, we found in Section \ref{sec:insta} that the Alfv\'{e}n speed vanishes on the firehose instability threshold. Therefore, Stage 2-firehose has,
\begin{equation}\label{eq:MA2S2f}
M_{A2x} = +\infty.
\end{equation}
As a consequence, when the system eventually settles in Stage 2-firehose with $\sigma_x < 1$ (i.e, $M_{A1x}>1$), it is evolutionary.

The anisotropy of Stage 2-firehose is also modified with respect to Paper I, due to the modified stability criterion. It now reads,
\begin{equation}
A_2 = 1-\frac{2 (T_2^3+T_2) \cos ^2\theta_1}{2 M_{A1}^2 (T_2-\tan \theta_1)+T_2^3 \cos ^2\theta_1+T_2 \sin ^2\theta_1+\sin 2 \theta_1}.
\end{equation}

\subsection{Stage 2 mirror}
Since the stability threshold for the mirror instability is the same as in Paper I, the polynomial for $T_2$, the anisotropy $A_2$ and the density ratio are also the same. The Alfv\'{e}n Mach number reads here
\begin{equation}\label{eq:MA2S2m}
M_{A2x}^2 = \frac{2}{3}\frac{\sec^2\theta_1}{r}M_{A1}^2 .
\end{equation}


\end{document}